\documentclass{agujournal2019}
\usepackage{url} 
\usepackage{lineno}
\usepackage[finalnew]{trackchanges} 
\usepackage{soul}
\usepackage{aas_macros}  
\usepackage{amssymb}  
\usepackage{amsmath}  
\let\oldequation\equation
\let\oldendequation\endequation
\renewenvironment{equation}
  {\linenomathNonumbers\oldequation}
  {\oldendequation\endlinenomath}

\draftfalse 

\journalname{JGR: Planets}

\begin{document}

%
%


\title{Storms and the Depletion of Ammonia in Jupiter: \\ II. Explaining the Juno observations}

%
%




\authors{Tristan Guillot$^{1,2}$, Cheng Li$^{3,4}$, Scott J. Bolton$^{5}$, Shannon T. Brown$^{6}$, Andrew P. Ingersoll$^{3}$, Michael A. Janssen$^{6}$, Steven M. Levin$^{6}$, Jonathan I. Lunine$^{7}$, Glenn S. Orton$^{6}$, Paul G. Steffes$^{8}$, David J. Stevenson$^{3}$}

\affiliation{1}{Universit\'e C\^ote d’Azur, OCA, Lagrange CNRS, 06304 Nice, France}
\affiliation{2}{The University of Tokyo, Department of Earth and Planetary Science, Tokyo 113-0033, Japan}
\affiliation{3}{California Institute of Technology, Pasadena, CA 91125, USA}
\affiliation{4}{University of California Berkeley, USA}
\affiliation{5}{Southwest Research Institute, San Antonio, Texas, USA}
\affiliation{6}{Jet Propulsion Laboratory,  California Institute of Technology, Pasadena, CA 91109, USA}
\affiliation{7}{Cornell University, Ithaca, New York 14853, USA}
\affiliation{8}{School of Electrical and Computer Engineering, Georgia Institute of Technology, Atlanta, GA, USA}





\correspondingauthor{Tristan Guillot}{tristan.guillot@oca.eu}




\begin{keypoints}
\item Juno measurements show that ammonia gas in Jupiter has variable abundance to great depth and as a function of latitude
\item We show that Jupiter's powerful storms control ammonia abundance by leading to the formation of water-ammonia hailstones (mushballs) and evaporative downdrafts
\item A simple atmospheric mixing model successfully links measured lightning rate to ammonia abundance and predicts variable water abundance to great depth. 
\end{keypoints}

%
%

%
%


\begin{abstract}
Observations of Jupiter's deep atmosphere by the Juno spacecraft have revealed several puzzling facts: The concentration of ammonia is variable down to pressures of tens of bars, and is strongly dependent on latitude. While most latitudes exhibit a low abundance, the Equatorial Zone of Jupiter has an abundance of ammonia that is high and nearly uniform with depth. In parallel, the Equatorial Zone is peculiar for its absence of lightning, which is otherwise prevalent most everywhere else on the planet. We show that a model accounting for the presence of small-scale convection and water storms originating in Jupiter’s deep atmosphere accounts for the observations. Where strong thunderstorms are observed on the planet, we estimate that the formation of ammonia-rich hail ('mushballs') and subsequent downdrafts can deplete efficiency the upper atmosphere of its ammonia and transport it efficiently to the deeper levels. In the Equatorial Zone, the absence of thunderstorms shows that this process is not occurring, implying that small-scale convection can maintain a near-homogeneity of this region. A simple model satisfying mass and energy balance accounts for the main features of Juno's MWR observations and successfully reproduces the inverse correlation seen between ammonia abundance and the lightning rate as function of latitude. We predict that in regions where ammonia is depleted, water should also be depleted to great depths. The fact that condensates are not well mixed by convection until far deeper than their condensation level has consequences for our understanding of Jupiter’s deep interior and of giant-planet atmospheres in general.
\end{abstract}

\section*{Plain Language Summary}
\remove{Measurements by the Juno spacecraft have shown that ammonia in Jupiter is present near the equator of the planet but is depleted to great depths at other latitudes, something never anticipated by theoretical models. In a companion paper, we showed that ammonia can combine to water to form hail-like particles (mushballs) that can fall to great depths. Here we show that storms can indeed effectively deplete the upper atmosphere to great depths. The dichotomy seen with the ammonia abundance between the equator and other regions is also seen in the measured flash rate, indicative of storm activity in Jupiter: No lightning has been detected at the equator in the region which has a high abundance of ammonia. We predict that water, another crucial species to understand Jupiter's meteorology and formation, is also depleted to great depths. Thus Jupiter's atmosphere is much more complex than anticipated, affecting how we understand its interior, composition and formation. This should also apply to other giant planets, and to exoplanets with hydrogen atmospheres.}

\add{Measurements by the Juno spacecraft have shown that much more ammonia is present in Jupiter's atmosphere near the equator than at higher latitudes. This was never predicted by theory.  In a companion paper, we showed that ammonia can combine with water to form hail-like particles that we call ``mushballs''. Here we show that mushball formation in storms can effectively dry out the atmosphere of its ammonia. Our idea is supported by lack of lightning activity at the equator compared to higher latitudes. Because lightning is generated in rainstorms, the lack of lightning at the equator suggests that the thunderstorms forming the mushballs responsible for the depletion of ammonia are not present there.  In contrast, in other regions where lightning is present, we predict that not only ammonia but also water are depleted to great depths, more than a hundred kilometers below the cloud tops. The complexity of Jupiter's meteorology means we must expect similar complexity in observing the weather on other giant planets in and beyond our solar system.}

%
%

%


%
%
%
%

\section{Introduction}

Jupiter is the archetype of planets with deep hydrogen atmospheres. Contrary to the Earth, it has no surface and all condensates are heavier than the main non-condensable constituants, hydrogen and helium. Recent observations reveal that its atmosphere is much more complex than traditionally assumed, with implications for its dynamics, the structure and internal composition of Jupiter and the evolution of planets with hydrogen atmospheres, including exoplanets. 

Jupiter is known for its alternance of dark reddish-zones and light, white belts. Besides their colors, these zones and belts are characterized by alternating zonal speeds that differ by up to about 100\,m/s \cite{GarciaMelendo+SanchezLavega2001,Porco+2003,Tollefson+2017}. But when observed at much longer wavelengths (1 to 60\,cm), the Juno microwave radiometer (MWR) sees a different structure: An equatorial region between latitudes $0^\circ$ and $5^\circ$N which is systematically \remove{colder}\add{darker} (lower brightness temperature) than all other latitudes and fainter variations between zones and belts \cite{Bolton+2017}. This reveals a puzzling dichotomy of Jupiter's deep atmosphere: In this $0^\circ-5^\circ$N latitudinal region, the atmosphere contains a high, vertically relatively uniform, abundance of ammonia whereas it is much lower and variable at other latitudes. The abundance of ammonia increases with depth and may become equal to the equatorial value, but at pressures of 30\,bars or more \cite{Li+2017}. 

Signs of the depletion of ammonia in Jupiter's atmosphere were obtained from ground-based radio-wave observations as early as 1986 \cite{dePater1986} \add{and the dichotomy between the equatorial region and other latitudes was discovered a few years later} \cite{dePater+2001, Showman+dePater2005, dePater+2019}, but the observations could not probe levels as deep as those accessible to Juno. \remove{The dichotomy between the equatorial region and other latitudes}\add{This dichotomy} is also seen in the 5-$\mu$m spectroscopic observations of Jupiter at 1-4 bar levels, although the retrieval is more complex due to the effects of clouds \cite{Giles+2017, Blain+2018}. 

\remove{This dichotomy}\add{Such a global change in ammonia abundance over the planet and down to great depths} cannot be explained solely by meridional circulation (\add{i.e., a Hadley-type circulation with} upward motion at the equator and downward motion at other latitudes) and requires a localized downward transport of ammonia that is essentially invisible to Juno's MWR instrument. \add{Indeed, if one considers large-scale advection only (i.e. assuming that ammonia rain is unimportant), satisfying both the observed ammonia distribution and the global mass-balance requires that the downdrafts are of {\it higher} ammonia concentration than the updrafts} \cite{Ingersoll+2017}. \add{An alternative hypothesis also involving a Hadley-type circulation would require efficient ammonia rainout in the upwelling equatorial branch, and compensating subsidence everywhere else. However, it is difficult to imagine how this extreme model could account for the planet's zones and belts }\cite{Fletcher+2020} \add{and it would be  at odds with the observation that storms are present at mid- and high-latitudes and not at the equator }\cite{Brown+2018}.
\add{As discussed by} \citeA{Ingersoll+2017}, we must therefore seek a process capable of (i) drying \add{out} the upper atmosphere of its ammonia to great depths, (ii) accounting for the dichotomy between the equatorial region and other latitudes while (iii) remaining sufficiently small-scale and/or intermittent to have escaped detection thus far. 

In a companion paper \cite{Guillot+2020a} (hereafter paper~I), we have shown that during strong storms able to loft water ice into a region located at pressures between 1.1 and 1.5 bar and temperatures between 173K and 188K, ammonia vapor can dissolve into water ice to form a low-temperature liquid phase containing about 1/3 ammonia and 2/3 water. \add{The presence of this liquid mixture is consistent with the observation of lightning flashes originating from low presssure levels} \cite{Becker+2020}. The subsequent formation of ammonia-rich hail that we call 'mushballs' leads to an effective transport of the ammonia to deep levels (between 7 and 25 bars, depending on poorly-known ventilation coefficients). Further sinking of ammonia- and water-rich plumes must take place because evaporation leads to a gas that has a high molecular weight and a low temperature due to evaporative cooling. 

This downward transport is a necessary but not sufficient condition to explain the observations: \remove{It can be argued that} Storms, particularly strong storms, cover a tiny fraction of the atmosphere of the planet and they are strongly intermittent. Based on our experience of Earth's storms, hail is rare (fortunately!). Lastly, mass balance implies that some of the ammonia-rich atmosphere from the deeper level must be transported upward. Given these observations how could hail (or mushball) formation be of significance in Jupiter? 

The present paper explores the consequences of the presence of mushballs and evaporative downdrafts for the atmosphere of Jupiter. Can such a process operate efficiently enough to yield a widespread depletion of ammonia in most of Jupiter's troposphere? Can it account for the main features of Juno/MWR measurements? What are its consequences for our understanding of Jupiter's atmospheric heat engine and for the distribution of water on the planet? We propose hereafter a simple local model to address these questions broadly, leaving aside for future work other important aspects like time-dependency and interplay between local vertical transport and global \remove{mixing}\add{atmospheric circulation}. 

The paper is organized as follows: In Section~\ref{sec:MWR}, we put the Juno MWR maps of inferred ammonia abundance in the context of a physical model of Jupiter's deep atmosphere. In Section~\ref{sec:model}, we then present a \remove{toy}\add{mass-exchange} model that solves mass- and energy-balance locally in Jupiter. We apply this model to interpret the MWR observations and derive consequences for our understanding of Jupiter's deep atmosphere in Section~\ref{sec:applications}.

\section{Juno's ammonia abundance map}\label{sec:MWR}

The Juno microwave radiometer measures the thermal radiation of Jupiter’s atmosphere at six radio wavelengths probing approximately from 0.7 to 250 bars. Because Jupiter is emitting more heat than it receives from the Sun \cite{Hanel+1981, Li+2018} and because radiative opacities are large \cite{Guillot+1994, Guillot+2004} it is believed that its deep atmosphere (below \add{$\sim 0.8$\,bar,} the ammonia condensation level) should be largely convective and adiabatic. This was confirmed within a few \remove{kelvins}\add{K} (see hereafter Section~\ref{sec:temp}) both by radio occultation from the Voyager spacecraft \cite{Lindal+1981} and in situ measurements of the Galileo probe \cite{Magalhaes+2002}. Assuming Jupiter’s temperature profile lies on an adiabat defined by the Galileo measurement (i.e., 166.1\,K at 1 bar), the variations of the brightness temperatures as a function of latitude and wavelength are entirely determined by the distribution of the ammonia gas, which is the major absorber in the wavelengths of Juno/MWR \cite{Janssen+2017}. The 2D distribution of ammonia is derived by fitting the microwave spectra at every latitude.  In \citeA{Li+2017}, the map was derived by using only the observation of the first perijove (PJ1). The subsequent observations probe different longitudes and are very similar to PJ1. Therefore, we use the average of the first 9 perijoves to produce the mean condition of Jupiter’s atmosphere across multiple longitudes. 

\begin{figure}
\noindent\includegraphics[width=\textwidth]{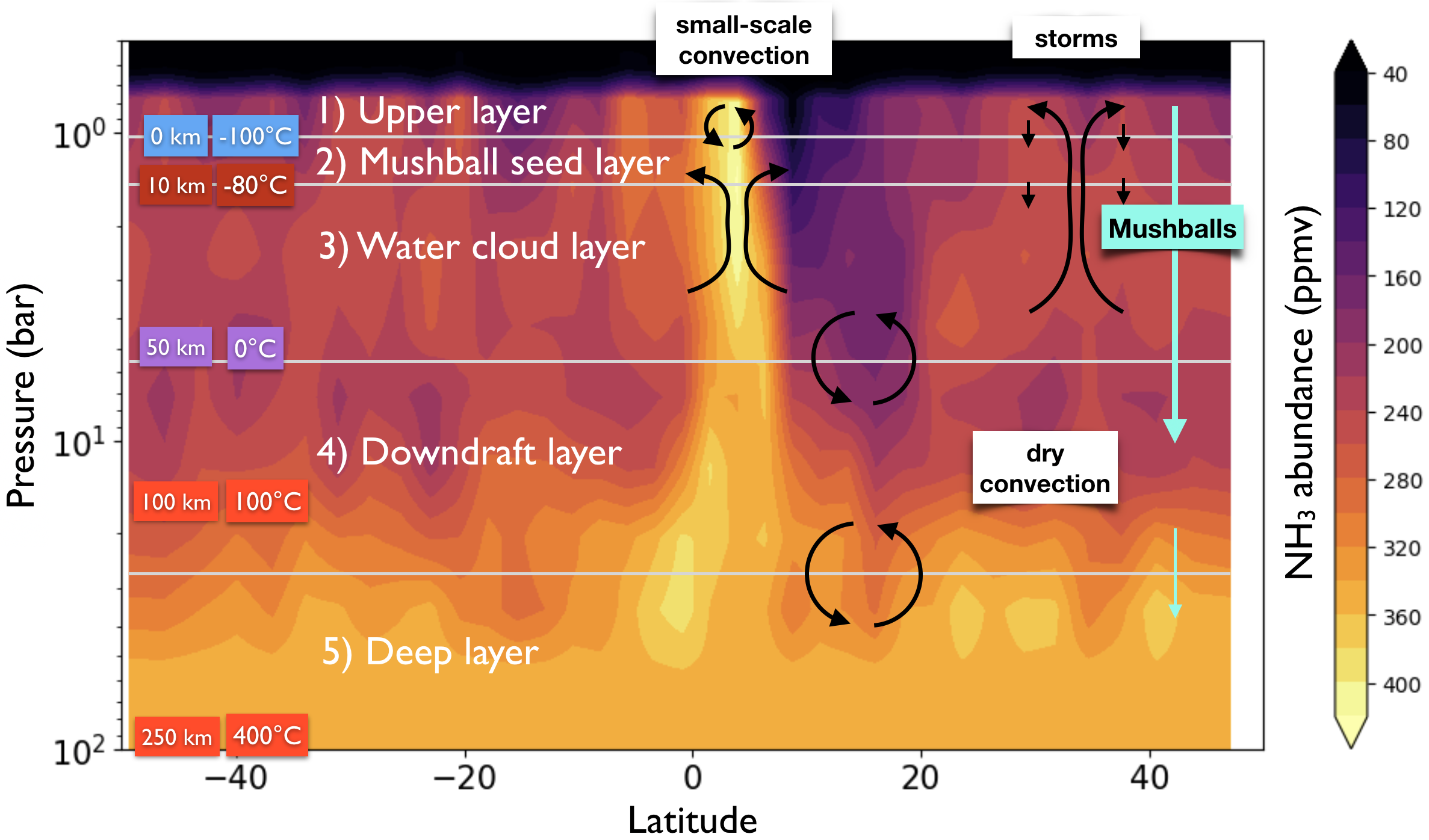}
\caption{Average map of ammonia abundance in Jupiter retrieved by the Juno MWR during PJ1 to PJ9 as a function of latitude and pressure. Overlaid are indications of altitude and temperature as well as the layers and mechanisms (small-scale convection and/or storms in the water condensation region, dry convection deeper) considered in this work (see text). Water vapor condenses to ice particles at $\sim 5$\,bar level ($0^\circ$C), $\sim 50$\,km below the 1\,bar level. }
\label{fig:model5layers}
\end{figure}

Figure~\ref{fig:model5layers} shows that for latitudes between $0^\circ$ and $5^\circ$N, the ammonia concentration is high, near its global maximum of $360\,$ppmv,  and mostly uniform with depth. (A small increase in the concentration above $360\,$ppmv near 1-3 bar may be reproduced by including the effect of ammonia rain \cite{Li+Chen2019, Li+2020}.)  Away from the equator, the atmosphere is depleted in ammonia from the higher levels, down to $\sim 30$\,bar or so, where it increases to its global maximum. A maximum depletion of ammonia is observed between latitudes $5^\circ$ and $20^\circ$N, with an abundance of order 100\,ppmv near 1\,bar increasing progressively to reach about 200ppmv near 10\,bar. Another local minimum with an ammonia abundance below 200\,ppmv is located between lat. $-12^\circ$ and $-18^\circ$S, but is limited to pressures smaller than 3\,bar. Aside from these regions, the ammonia abundance below 10 bars fluctuates with altitude between 200 and 250\,ppmv and rises progressively to about 360\,ppmv at pressures between 30 and 100\,bar. 

These features are shared on all the passes observed with MWR and are thus very stable (an exception is the location of the Great Red Spot, which we do not consider here). There are fluctuations from one pass to the next but they are limited in magnitude and in range. In particular, the Equatorial Zone between $0^\circ$ and $5^\circ$N always shows a high nearly uniform abundance of ammonia near 360\,ppmv, the region between $5^\circ$ and $20^\circ$ is always the most depleted down to about 10-20 bar and the second minimum at pressures smaller than about 3 bar is always near $-16^\circ$. \add{The MWR measurements uncertainties are estimated to be of order 30\,ppmv. They are dominated by an absolute uncertainty on the calibration of order 2\%. The relative uncertainty (between different locations) should be significantly smaller} \cite{Li+2017}.

For the deeper levels, the information in Fig.~\ref{fig:model5layers} relies on data from MWR channels 2 and 1 whose weighting functions are very broad and peak around 30\,bar and 250\,bar, respectively \cite{Janssen+2017}.  This implies that the pressure at which the ammonia abundance starts rising (i.e., 20 bars or so) is uncertain. Also, we cannot distinguish between a progressive or sudden change.

\section{A \remove{toy}\add{mass-exchange} model for Jupiter's atmosphere}\label{sec:model}

We now develop a simple, \remove{toy}\add{mass-exchange} model of Jupiter's deep atmosphere. We choose \remove{an extreme}\add{a simple} approach, namely to assume that horizontal mixing \remove{may be neglected so that a steady-state may be achieved at each latitude/longitude in Jupiter} \add{takes place on longer timescales than vertical mixing. For example, a simplified analysis based on the measured eddy velocity covariance at the cloud level indicates that it may take 3 years for a parcel of air to move from latitude $0^\circ$ to latitude $\pm5^\circ$} \cite{Ingersoll+2017}\add{, a time-scale that may be longer than vertical adjustments. We also neglect time-variability to look for the steady-state solution at each  latitude/longitude in Jupiter}.  We first describe \add{the model principle}, derive its governing equations, find some analytical solutions and show how the ammonia abundance, water abundance and potential temperature vary as a function of the frequency of water storms. 

\subsection{\add{Model principle}}
In order to test whether the formation of mushballs can reproduce the basic features of the Juno MWR map in Fig.~\ref{fig:model5layers}, we build a simple, 5-layer model based on the properties of the different regions.  From top to bottom, these layers are: (1) the upper atmosphere, (2) the mushball-forming region, (3) the water-cloud region, (4) the downdraft region and (5) the deep interior. Ammonia vapor is present in all regions, but water vapor is present only in layers (3), (4) and (5) (it is present as ice in regions (1) and (2) but only intermittently). 

We furthermore consider that transport in the water condensation region (layers 3 to 1) can occur either through small-scale convection \add{(i.e., convection events not primarily driven by latent heat release and occurring on a scale equivalent to a pressure scale height or less)} or through large water storms \add{(i.e., plumes driven by latent heat release and with a large vertical extent, from the base of the water cloud near 6 bars to the top of the tropopause at pressures below 1 bar)}. In the deeper interior, from layers 5 to 3, transport of interior heat and chemical species is done by dry convection. We expect small-scale convection to occur when moist convection is inhibited (e.g. because of mass loading or vertical shear). Small-scale\ convection is expected to transport elements and heat across adjacent layers. Rain or snow may occur but without any transport of the condensates across the different layers.  Thunderstorms should occur in the water-cloud region (3) whenever conditions are favorable (moist convection is not inhibited). We envision that they should lead to an upward transport of ice particles through the mushball-forming region (2) and into the upper region (1). 

On the basis of the observation of a large complex of storms in Jupiter’s atmosphere by the Galileo mission \cite{Gierasch+2000}, we envision that large storms should be the dominant mode of heat transport between the water cloud base (3) and the top layer (1). The frequency of these storms could be defined by the radiative timescale and the requirement to build convective available potential energy (CAPE) in order to exceed the buoyancy threshold \cite{Guillot1995, Li+Ingersoll2015}. At deeper levels, dry convection should occur, possibly powered by deeper ``rock storms''  created by the condensation of silicates and iron \cite{Markham+Stevenson2018}. 

Mushballs may form only when ice particles are transported to level (2) (Fig. 3), i.e. during thunderstorm events. Once formed, we envision that they rain down below the water-cloud base, to region (4) where they vaporize and partially to region (5) through downdrafts. The mean location of these five layers is set to $P_1=1$\,bar, $P_2=1.3$\,bar, $P_3=4$\,bar, $P_4=8$\,bar and $P_5=20$\,bar. While the location of the first three layers are set by physical and thermodynamical constraints (the properties of the upper atmosphere, the location of the mushball-formation and water-condensation regions), we note that the average pressures for layers 4 and 5 are loosely guided by the MWR results \add{and} largely unconstrained at this point.

\subsection{Governing equations}

Let us consider mass and energy balance in our simple 5-layer model shown in Fig.~\ref{fig:model5layers}. We define as \textit{c}\textsubscript{1},$ \ldots $ ,\textit{c}\textsubscript{5} the abundances of NH\textsubscript{3} in the 5 layers, \textit{w}\textsubscript{1},$ \ldots $ ,\textit{w}\textsubscript{5} the abundances of H\textsubscript{2}O (with \textit{w}\textsubscript{1}= \textit{w}\textsubscript{2}=0) and $ T_1 $,$ \ldots $ ,$ T_5 $ their temperatures. We \remove{fix}\add{prescribe} the bulk (bottom) mixing ratios of NH\textsubscript{3}, $c$, and water, $w$, and impose that the atmosphere must transport a known internal heat flux  \( F_{\rm tot} \)  \cite{Li+2018}. \add{The parameters of our mass-exchange model are summarized in} Table~\ref{tab:parameters}. 

\begin{table}
\caption{Parameters of our global model}\label{tab:parameters}
{\small
\begin{tabular}{lll}
\hline
Variable & Note & Fiducial value \\
\hline
$c$ & Bulk mass mixing ratio of ammonia & 0.0027 \\
$w$ & Bulk mass mixing ratio of water & 0.021 \\
$F_{\rm tot}$ & Internal heat flux & \\
$\dot{m}_{\rm conv}$ & Upward convective mass flux (layers $2\leftrightarrow 1$ \& $3\leftrightarrow 2$ )  & \\
$\dot{m}_{\rm storm}$ & Upward mass flux due to water storms (layers $3\rightarrow 1$) & \\
$\dot{m}_{\rm deep}$ & Upward convective deep mass flux (layers $4\leftrightarrow3$ \& $5\leftrightarrow 4$) &  \\
$a_{\rm w}$ & Fraction of water in mushballs ending in layer 4 & 0.5 \\
$\epsilon$ & Efficiency of mushball formation & 0.3 \\
$f_{\rm NH_3}$ &  Fraction of NH$_3$ in mushballs & 0.1 \\
$f_{\rm H_2O}$ &  Fraction of H$_2$O in mushballs & 0.9 \\
$q_{\rm mush}$ & Mass mixing ratio of condensables in downward plumes from levels 1 to 4 & 1 \\
$q_{\rm down}$ & Mass mixing ratio of condensables in downward plumes from levels 4 to 5 & $2w$ \\
\\
$c_1$ to $c_5$ & Ammonia mass mixing ratio in layers 1 to 5 &  \\
$w_1$ to $w_5$ & Water mass mixing ratio in layers 1 to 5 &  \\
$s_1$ to $s_5$ & Dry static stability in layers 1 to 5 &  \\
$\theta_1$ to $\theta_5$ & Potential temperature in layers 1 to 5 &  \\
$M_1$ to $M_5$ & Masses of layers 1 to 5 & \\
$P_1$ to $P_5$ & Average pressures of layers 1 to 5 & \\
$c_{\rm mush}$ & Surface-average mixing ratio of ammonia in sinking mushballs & \\
$w_{\rm mush}$ & Surface-average mixing ratio of water in sinking mushballs & \\
$\varpi_{\rm mush}$ & See eq.~\ref{eq:varpi_mush} & \\
$\varpi_{\rm down}$ & See eq.~\ref{eq:varpi_down} & \\
$f_{\rm storm}$ & $\equiv \dot{m}_{\rm storm} /\dot{m}_{\rm deep}$ & \\
$f_{\rm conv}$ & $\equiv \dot{m}_{\rm conv} /\dot{m}_{\rm deep}$ & \\
$L_{\rm v}$ & Latent heat of vaporization of water (at $0^\circ$C) & $2.52\times 10^{10}$\,erg/g \\
\hline
\end{tabular}
}
\end{table}

\begin{figure}
\includegraphics[width=13cm]{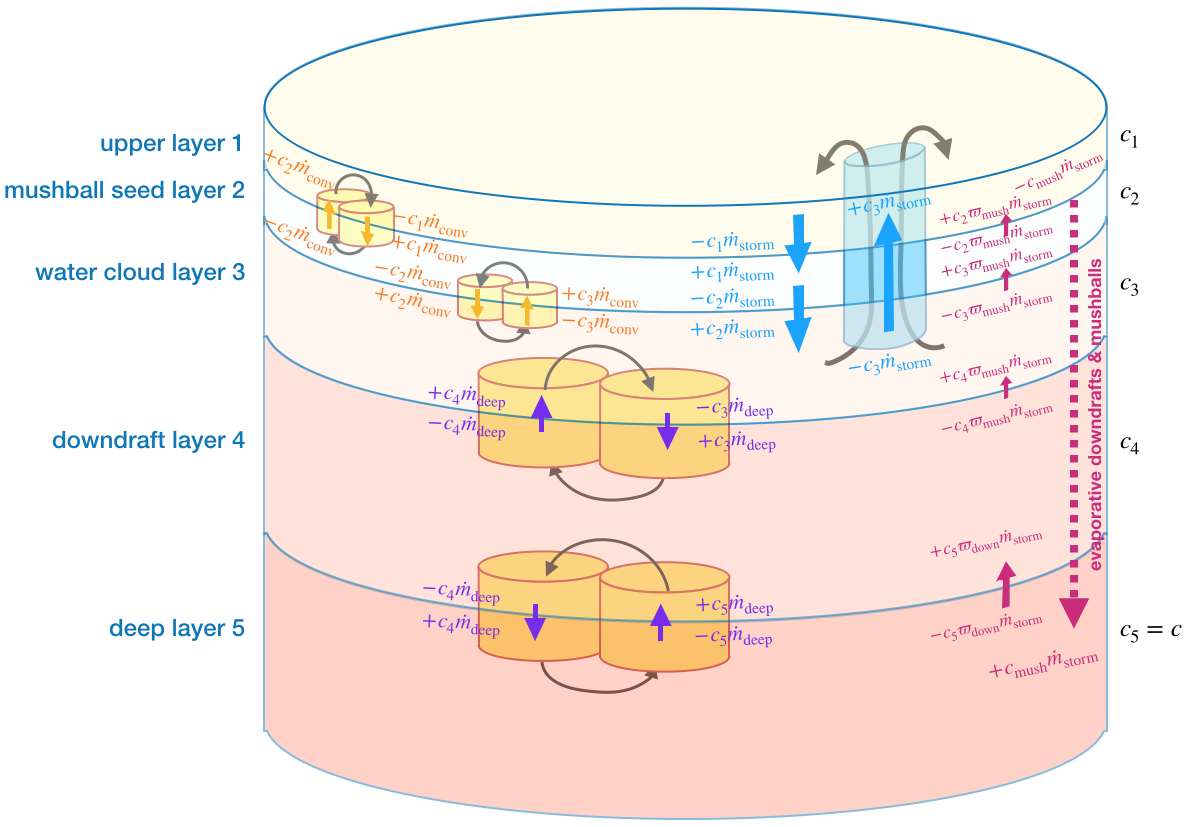}
\caption{Mass balance of ammonia in the framework of our 5-layer model. We consider that three main processes transport material between layers: In yellow, small-scale convection is modeled as an updraft and its reciprocal downdraft between adjacent layers. We consider that it is characterized by an upward mass flux $\dot{m}_{\rm deep}$ between layers 5 and 4 and layers 4 and 3, and by an upward mass flux $\dot{m}_{\rm conv}$ between layers 3 and 2 and layers 2 and 1. In blue, strong storms due to water condensation lead to a transport of material directly from layer 3 to layer 1 and to a compensating subsidence mass flux from layer 1 to layer 2 and to layer 3. These storms also lead to the formation of mushballs and evaporative downdrafts which deliver ammonia and water directly to layers 4 and 5. The terms in each layer correspond to the mass balance of ammonia described by Eq.~(\ref{eq:5layers}).}
\label{fig:col_nh3}
\end{figure}

\begin{figure}
\includegraphics[width=13cm]{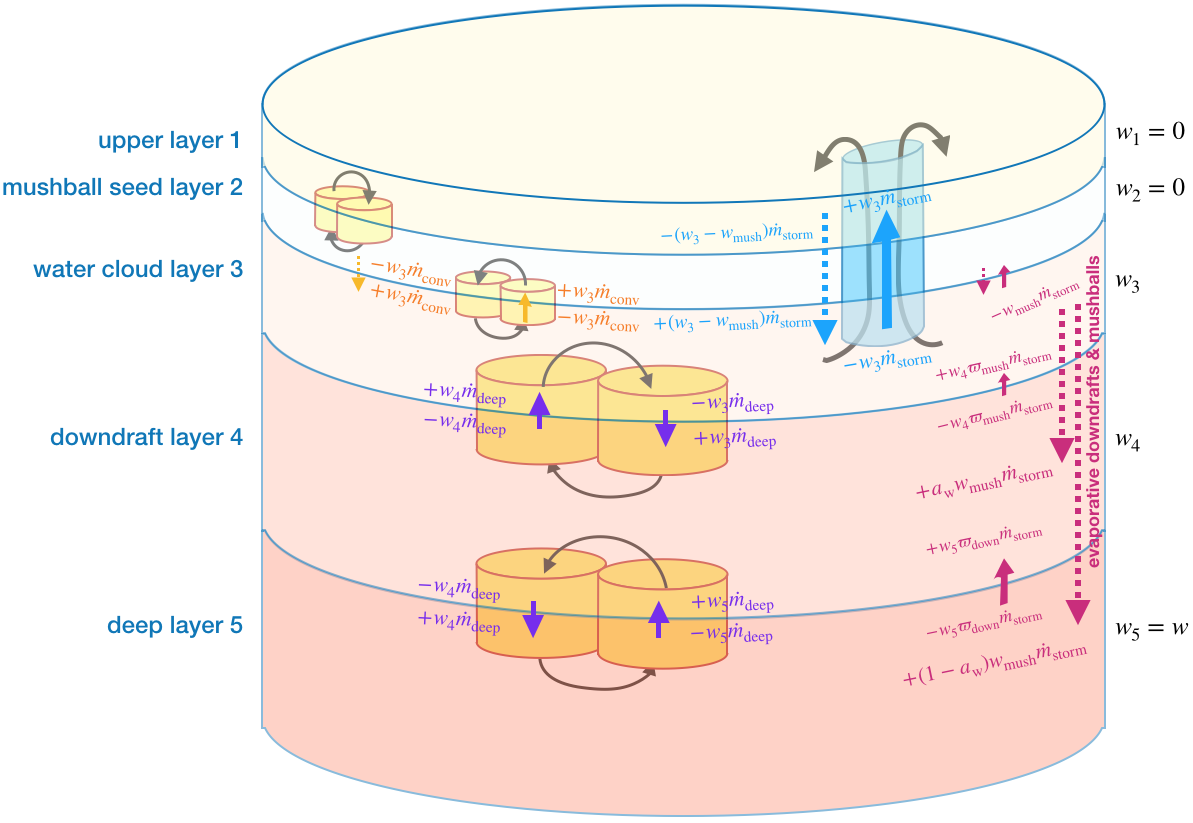}
\caption{As Fig.~\ref{fig:col_nh3} for the mass balance of water in the framework of our 5-layer model. \add{Compared to} Fig.~\ref{fig:col_nh3}, \add{the main differences arise from the fact that water precipitates out of layers 2 and 1 so that its abundance in these layers is, on average, extremely small. Also, the evaporation of mushballs removes a fraction $a_{\rm w}$ of the water which is incorporated into layer 4, with the remaining $1-a_{\rm w}$ being incorporated into layer 5.} }
\label{fig:col_h2o}
\end{figure}

We consider storms and convective mixing as discrete events connecting the different layers. Our approach including all the terms included to calculate the mass balance of ammonia and water is shown hereafter in Figs.~\ref{fig:col_nh3} and \ref{fig:col_h2o}, respectively. The three mechanisms that we envision lead to an upward transport of material per unit time $\delta t$ of a mass $\dot{m}_{\rm conv}\delta t$, $\dot{m}_{\rm storm}\delta t$ and $\dot{m}_{\rm deep}\delta t$, respectively. The same mass is also transported downward either as part of the downward convective cell or due to compensating subsidence. 

In addition, on the basis of the findings of Paper~I, we envision that a downward flux of mushballs deliver a mass of ammonia $c_{\rm mush}\dot{m}_{\rm storm}\delta t$ down to layer 5, and a mass of water that is split between $a_{\rm w}w_{\rm mush}\dot{m}_{\rm storm}\delta t$ to layer 4 and $(1-a_{\rm w})w_{\rm mush}\dot{m}_{\rm storm}\delta t$ to layer 5, with $a_{\rm w}$ being a parameter between 0 and 1. 


The mushball mass flux is parameterized as follows: We consider that the mushball efficiency mechanism is proportional to the difference between the mixing ratio in layer 2 and the minimum mixing ratio for the process to operate,  $\approx 100\,$ppmv (see Paper~I). We also consider that the mushball flux is limited by the amount of water present in the water cloud layer,  \( w_{3} \) . The flux itself is proportional to the mass flux due to storms. We thus write:
\begin{equation}
\left\{
\begin{array}{l}
 c_{\rm mush}= \epsilon  \min\left( c_{2}-c_{\rm min},w_{3}{f_{\rm NH_3}}/{f_{\rm H_2O}} \right) ,\\
 w_{\rm mush}= c_{\rm mush}{f_{\rm H_2O}}/{f_{\rm NH_3}} ,
\end{array}
\right.
\end{equation}
where  \(  \epsilon  \)  is an efficiency parameter ( \( 0 \leq  \epsilon  \leq 1 \) ) \add{which corresponds to the fraction of ammonia in the storm that is eventually embedded in the mushballs},  and $f_{\rm NH_3}$ and $f_{\rm H_2O}$ are the mass mixing ratios of condensed ammonia and water in the mushballs, respectively. Our fiducial parameters based on our simple mushball evolution model \add{from Paper~I} are  \(  \epsilon  \) =0.3,  \( f_{\rm NH_3} \) =0.1 (thus  \( f_{\rm H_2O} \) =0.9),  \( a \) =0.5. 

The total downward mushball flux to level 4 is thus
\begin{equation}
\dot{\tilde{m}}_{\rm mush, 1\rightarrow 4}=(c_{\rm mush}+w_{\rm mush}) \dot{m}_{\rm storm}=(w_{\rm mush}/f_{\rm H_2O}) \dot{m}_{\rm storm}.
\end{equation}
Where the~''$\;\tilde{ }\;$''~sign indicates that only condensates are considered. In addition, some air may be entrained down with the mushballs. Let us define $q_{\rm mush}$, the mass fraction of mushballs in that downward stream. The upward flux to compensate for the flux of mushballs and entrained air is thus: 
\begin{equation}
  \dot{m}_{1\rightarrow 4}=\dot{\tilde{m}}_{\rm mush, 1\rightarrow 4}/q_{\rm mush} \equiv \varpi_{\rm mush}\dot{m}_{\rm storm},
\label{eq:varpi_mush}
\end{equation}
where $\varpi_{\rm mush}=w_{\rm mush}/(f_{\rm H_2O}q_{\rm mush})$. We will assume that until mushballs evaporate, the fraction of air that is entrained is small, hence $q_{\rm mush}\approx 1$. 

Between level 4 and level 5 we consider that part of the mushballs have been stripped of their water and that even after full evaporation further sinking proceeds because of downdrafts powered by evaporative cooling (see Paper~I). The downward flux of ammonia is thus $c_{\rm mush}\dot{m}_{\rm storm}$ and the downward flux of water $(1-a_{\rm w}) w_{\rm mush}\dot{m}_{\rm storm}$. Thus, the total downward flux of condensates is 
\begin{equation}
\dot{\tilde{m}}_{\rm mush, 4\rightarrow 5}=(c_{\rm mush}+(1-a_{\rm w})w_{\rm mush}) \dot{m}_{\rm storm}=w_{\rm mush}\left(\frac{1}{f_{\rm H_2O}}-a_{\rm w}\right) \dot{m}_{\rm storm}.
\end{equation}
As previously, we account for the entrainment of air in the downdraft, with a mass fraction of condensates $q_{\rm down}$. This time, two limiting cases are $q_{\rm down}\sim 1$ if mushballs do reach layer 5 before evaporating (e.g., if ventilation coefficients have been overestimated -- see Paper~I), and \add{$q_{\rm down}\sim 0$} otherwise. As previously, the compensating upward flux is
\begin{equation}
  \dot{m}_{4\rightarrow 5}=\dot{\tilde{m}}_{\rm mush, 4\rightarrow 5}/q_{\rm mush} \equiv \varpi_{\rm down}\dot{m}_{\rm storm},
\label{eq:varpi_down}
\end{equation}
where $\varpi_{\rm down}=\left(1-a_{\rm w}f_{\rm H_2O}\right) w_{\rm mush}/(f_{\rm H_2O}q_{\rm down})$. 

Let us consider as an example layer 1, of mass $M_1$ and ammonia mixing ratio $c_1$. As shown in Fig. ~\ref{fig:col_nh3}, small-scale convection brings per time $\delta t$ a mass of ammonia $c_2\dot{m}_{\rm conv} \delta t$ and removes $c_1\dot{m}_{\rm conv} \delta t$. Similarly, storms deliver directly from layer 3 to layer 1 a mass of ammonia $c_3\dot{m}_{\rm storm} \delta t$ and compensating subsidence removes at the same time a mass $c_1\dot{m}_{\rm storm} \delta t$. These storms also lead, through the formation of mushballs, to a removal of $c_{\rm mush}\dot{m}_{\rm storm} \delta t$ of ammonia, which is transported directly to layer 5 and to a compensating upward mass flux of ammonia $c_2 \dot{m}_{\rm mush}\delta t$. Thus, the change in ammonia mass in layer 1 is 
\begin{equation}
\delta c_1 M_1=(c_2-c_1) \dot{m}_{\rm conv} \delta t + (c_3-c_1-c_{\rm mush}+c_2\varpi_{\rm mush})\dot{m}_{\rm storm} \delta t. \nonumber
\end{equation}
 Since we are looking for a steady-state solution, the equation governing the ammonia mass balance for layer 1 is 
\begin{equation}
 0=(c_2-c_1) \dot{m}_{\rm conv} + (c_3-c_1-c_{\rm mush}+c_2\varpi_{\rm mush})\dot{m}_{\rm storm},  \nonumber
\end{equation}
i.e., a simple equation independent of the mass of the layer itself. The same approach can then be used for each layer. In order to close the system, we choose as limiting condition that the mixing ratio of the bottom layer is prescribed to the value inferred from the Juno measurement. 


For water, with a mixing ratio $w$, the equations are the same, but we must consider that water is only present in condensed form in layers 1 and 2 and will therefore very rapidly be transported back to layer 3. Also, on the basis of Paper~I, we consider that a fraction $a_{\rm w}$ of the mushballs are evaporated in level~4 and its counterpart $(1-a_{\rm w})$ in level~5. Only 3 equations are needed for level 3 and 4 and to close the system with $w_5=w$. The resulting mass balance is represented in Fig.~\ref{fig:col_h2o}. Since layers 1 and 2 have a median abundance of water that is negligible, only 3 equations are needed for level 3 and 4 and to close the system with $w_5=w$. As an example, the mass balance equation for water in layer 3 is: 
\begin{equation}
\delta w_3 M_3=(w_4-w_3) \dot{m}_{\rm deep} \delta t + (w_5-w_3-w_{\rm mush}+w_4\varpi_{\rm down})\dot{m}_{\rm storm} \delta t.  \nonumber
\end{equation}
As for ammonia, the steady-state solution \add{($\delta w_3=0$)} is independent of layer mass. 

\begin{figure}
\includegraphics[width=13cm]{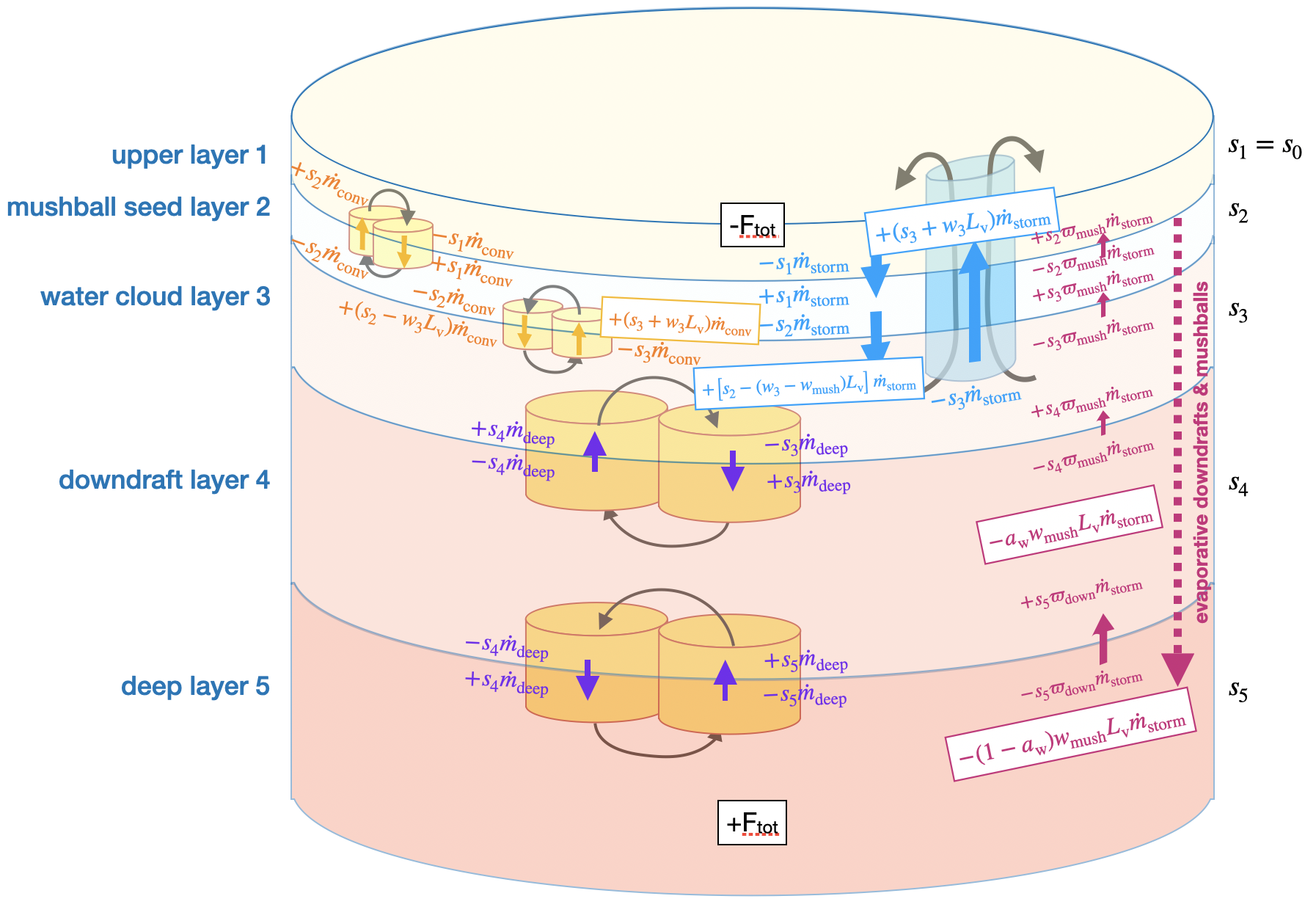}
\caption{As Fig.~\ref{fig:col_nh3} for the balance of static energy in the framework of our 5-layer model. In addition to the terms due to a transport of static energy, terms resulting from the condensation or evaporation of water are highlighted. \add{A flux $F_{\rm tot}$ arising from internal heat is added to layer 5 and removed from layer 1.}}
\label{fig:col_s}
\end{figure}

Finally we consider in Fig.~\ref{fig:col_s} energy balance in the system. Since we consider levels at relatively high optical depth, we neglect any radiation heating/cooling. Dry static energy, $s\equiv c_P T+gz$ with $c_P$ being the heat capacity of air and $z$ altitude, is therefore conserved during dry adiabatic motions. When condensation occurs in updrafts or due to evaporation, moist static energy $h=c_P T + gz+ L_{\rm v} w$  with  \( L_{\rm v} \)  being the latent heat of vaporization of water, is approximately conserved \cite{Holton1992}. (For this simple model, we neglect the effect of the condensation of ammonia because of its expected much smaller abundance). Equivalently, dry static energy is increased by $L_{\rm v} w$ by the condensation of water, or decreased by the same amount upon vaporization. 

As illustrated by Fig.~\ref{fig:col_s}, dry convective events result in mixing static energy between adjacent layers. Small-scale convection results in condensation of transported water in layer 2 and its vaporization in layer 3, resulting in positive and negative contributions in these respective layers. Storms lead to condensation of water and transport of the static energy to level 1. Part of the water flux is reevaporated in layer 3. The other part forms mushballs which reevaporate (and deliver a negative static energy contribution) in layers 4 and 5. Note that in this simple model, we do not consider the small contribution of water (or ammonia) gases to the static energy budget and we also neglect any possible condensation events linked to the \remove{small} upward mass flux that balances the downward flux of mushballs. 

As an example, for layer 2, we must consider the advection of static energy to and from adjacent layers, and we have to include a term due to the release of latent heat due to water condensation during small-scale convection events. Thus, 
\begin{equation}
\delta s_{2} M_2=(s_{3}+s_{1}-2s_{2} +w_3 L_{\rm H_2O}) \dot{m}_{\rm conv} \delta t + (s_{1}-s_{2})\dot{m}_{\rm storm} \delta t.  \nonumber
\end{equation}
For layer 5, we have to consider the internal heat flux $F_{\rm tot}\delta t$. Accounting for the evaporation of mushballs and static energy transport the energy budget for that layer is:
\begin{equation}
\delta s_{5} M_5=(s_{4}-s_{5}) \dot{m}_{\rm conv} \delta t + \left[-(1-a_{\rm w})w_{\rm mush}L_{\rm v}-s_5\varpi_{\rm deep} \right]\dot{m}_{\rm storm} \delta t+F_{\rm tot}\delta t.  \nonumber
\end{equation}

Overall, because we are looking for a steady-state solution \add{($\delta c_1=...=\delta s_5=0$)}, the solution is independent of the value of the mass flux \remove{itself,}\add{$\dot{m}_{\rm deep}$}. It is convenient to define
\begin{equation}
\left\{
\begin{array}{l}
f_{\rm conv}\equiv \dot{m}_{\rm conv} /\dot{m}_{\rm deep}\\
f_{\rm storm}\equiv \dot{m}_{\rm storm} /\dot{m}_{\rm deep}\\
\end{array}
\right.
\end{equation}
\add{The value of $f_{\rm conv}+f_{\rm storm}$ is thus a measure of how much mass is concerned by convective motions per unit time in layers 1-3 (i.e., $\dot{m}_{\rm conv}+\dot{m}_{\rm storm}$)  compared to the same value in layers 4-5 ($\dot{m}_{\rm deep}$). }

We can thus obtain 5 equations for the ammonia mass balance, 3 for the water mass balance and 5 for the energy balance (including the 3 boundary conditions), as follows: 
\begin{equation}
\left\{
\begin{array}{l}
 \left( c_{2}-c_{1} \right) f_{\rm conv}+ \left[ c_{3}-c_{1}-c_{\rm mush}+c_2\varpi_{\rm mush} \right] f_{\rm storm}=0,\\
 \left( c_{3}+c_{1}-2c_{2} \right) f_{\rm conv}+ \left[ c_{1}-c_{2} +(c_3-c_2)\varpi_{\rm mush}\right] f_{\rm storm}=0,\\
 \left( c_{4}-c_{3} \right) + \left( c_{2}-c_{3} \right) f_{\rm conv}+ \left[ c_{2}-c_{3} +(c_4-c_3)\varpi_{\rm mush} \right] f_{\rm storm}=0,\\
 \left( c_{5}+c_{3}-2c_{4} \right)+\left[ c_5\varpi_{\rm down}-c_4\varpi_{\rm mush}\right] f_{\rm storm} =0,\\
c_{5}=c,\\
 \left( w_{4}-w_{3} \right) +\left[ -w_{\rm mush}+w_4\varpi_{\rm mush}\right] f_{\rm storm}=0\\
 \left( w_{5}+w_{3}-2w_{4} \right) +\left[ a_{\rm w}w_{\rm mush}+w_5\varpi_{\rm down}-w_4\varpi_{\rm mush}\right] f_{\rm storm}=0,\\
w_{5}=w,\\
 s_{1}=s_0,\\
\left(  s_{3}+ s_{1}-2 s_{2} +w_{3}L_{\rm v} \right)  f_{\rm conv}+ \left[  s_{1}- s_{2} + (s_3-s_2) \varpi_{\rm mush} \right] f_{\rm storm}=0,\\
\left(  s_{4}- s_{3} \right) + \left(  s_{2}- s_{3} -w_{3}L_{\rm v} \right)  f_{\rm conv}+ \left[  s_{2}- s_{3} - \left( w_{3}-w_{\rm mush} \right) L_{\rm v} + (s_4-s_3) \varpi_{\rm mush}  \right] f_{\rm storm}=0,\\
\left(  s_{5}+ s_{3}-2 s_{4} \right) +\left[ -a_{\rm w}w_{\rm mush}L_{\rm v} +s_5\varpi_{\rm down}-s_4\varpi_{\rm mush} \right] f_{\rm storm}=0,\\
\left(  s_{4}- s_{5} \right) +\left[  -\left( 1-a_{\rm w} \right) w_{\rm mush}L_{\rm v} -s_5\varpi_{\rm down} \right] f_{\rm storm}+F_{\rm tot}/\dot{m}_{\rm deep}=0.
\end{array}
\right.
\label{eq:5layers}
\end{equation}

\subsection{Static energy and potential temperature}\label{sec:temp}

\remove{Instead of}\add{As an alternative to} static energy, it is generally convenient to express the results in terms of potential temperature
\begin{equation}
\theta\equiv T (P/P_0)^{-{\cal R}/c_P},
\label{eq:theta}
\end{equation}
For a dry atmosphere and a perfect gas, the potential temperature defined by Eq.~(\ref{eq:theta}) \remove{by} is directly linked to the entropy. For a real atmosphere, the changes in specific heat, mean molecular weight and the departures from an ideal gas are thought of being relatively small (at the percent level), so that the potential temperature at deep levels can be used as a useful estimate of the boundary condition that should be used for interior models. Current interior models are generally based on the Voyager measurements of $165\pm 5 \,$K at 1\,bar \cite{Lindal1992, Guillot2005}. The Galileo probe measured a temperature at 1\,bar of $166.1\pm 0.2$\,K \cite{Seiff+1998}. For a dry adiabatic atmosphere \add{and setting $P_0$=1\,bar,} we would thus expect that at deep levels in Jupiter $\theta \approx 166\,$K. However the Galileo probe measured a temperature at 22\,bar of $427.7\pm 1.5\,$K \cite{Seiff+1998}, about 4\,K colder than expected for a dry adiabat \cite{Leconte+2017}. Assuming ${\cal R}/c_P \sim 0.3$, this implies a change in potential temperature $\Delta \theta\sim -1.6$\,K. 

In order to link the deviations in static energy to those in potential temperature in our simple model, we use the fact that $ds=c_P dT+ g dz=c_PT d\theta/\theta$. Using Eq.~\ref{eq:theta}, this implies
\begin{equation}
d\theta=\left(\frac{P}{P_0}\right)^{-{\cal R}/c_P}\frac{ds}{c_P},
\end{equation}
i.e. the deviations of the potential temperature at each level can be obtained by integrating changes in the static energy at each level. 

We thus derive the potential temperature difference at 1 bar as $\Delta\theta_i=\theta_i-\theta_1$ based on the static energies for each level calculated from Eq.~\ref{eq:5layers}, the pressure levels defined in Section~\ref{sec:MWR} and ${\cal R}/c_P=0.3$.

\subsection{Solutions as a function of $f_{\rm storm}$}

\begin{figure}
\noindent\includegraphics[width=12cm]{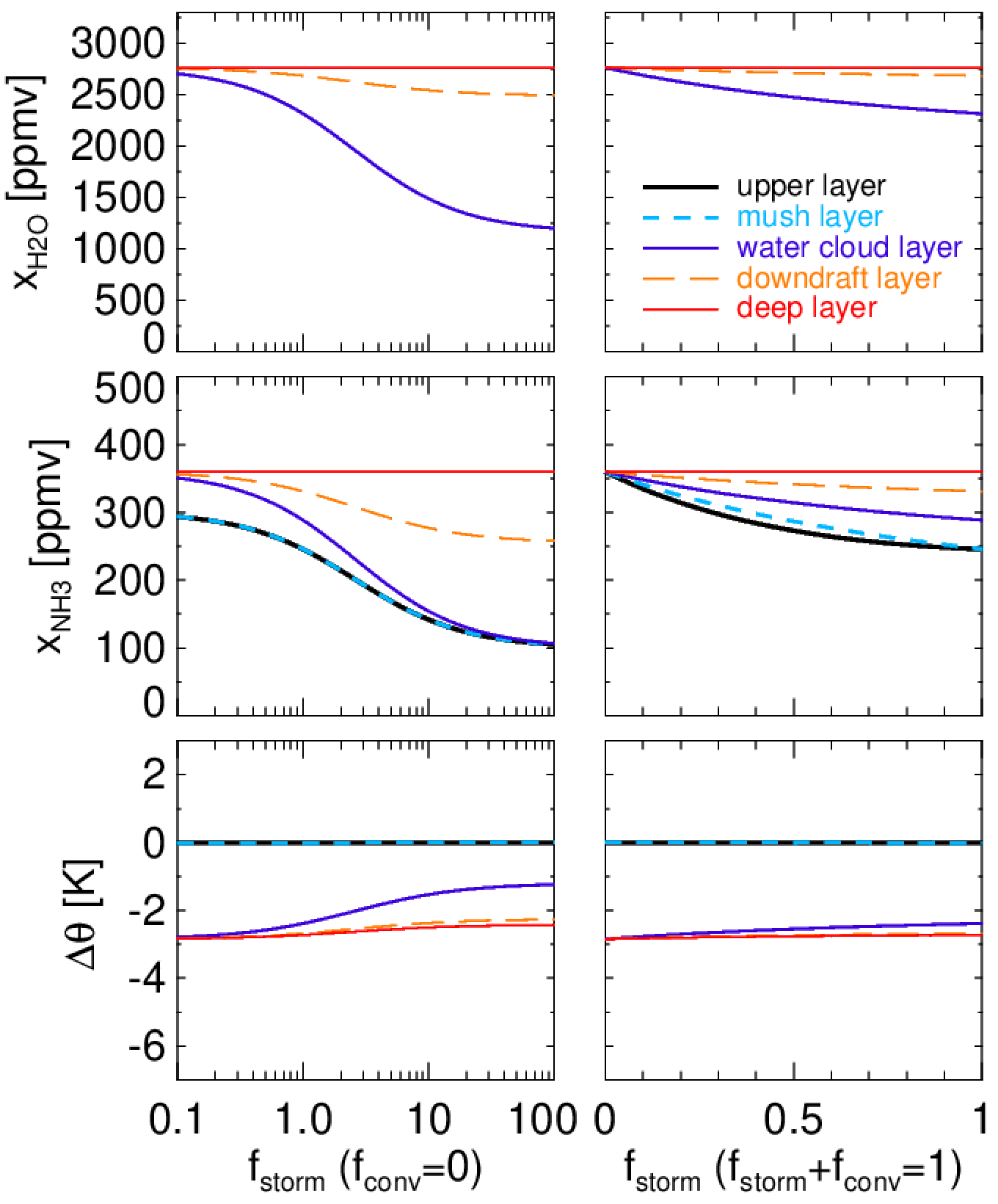}
\caption{Abundances of water (top row), ammonia (middle row) and potential-temperature anomalies (bottom row) obtained with our model, as a function of $f_{\rm storm}$, a parameter assessing the mass flux in large water storms relative to that of dry convection below the water cloud base. The left column corresponds to a situation in which no small-scale convection is present in the water condensation region ($f_{\rm conv}=0$) and pertains to mid-latitude regions of Jupiter. The right column assumes that both small-scale convection and storms occur, so that $f_{\rm storm}+f_{\rm conv}=1$.  The curves show the different layers considered in Fig.~\ref{fig:mush_model}: 1) upper layer (purple); 2) mushball-seed layer (blue, dashed); 3) water cloud layer (light blue); 4) downdraft layer (orange, dashed); 5) deep (red). The potential-temperature anomalies are calculated assuming that intrinsic heat flux transport occurs with negligible superadiabaticity (see text). }
\label{fig:mush_model}
\end{figure}

We now examine the solutions of Eq.~\ref{eq:5layers} as a function of our $f_{\rm storm}$ parameter for our fiducial parameters (see Table~\ref{tab:parameters}). Figure~\ref{fig:mush_model} shows the resulting mixing ratios of H$_2$O and NH$_3$ and the potential-temperature anomalies for the 5 layers considered. For convenience, we plot the solutions in terms of the volume mixing ratios, calculated with the approximate relations $x_{\rm NH_3}\approx (\mu/\mu_{\rm NH_3})c$ and $x_{\rm H_2O} \approx (\mu/\mu_{\rm H_2O})w$.

The two columns of Fig.~\ref{fig:mush_model} correspond to two different situations. The left column corresponds to a case in which storms carry most of the internal heat in the water condensation region, a situation that is relevant to the mid-latitudes in Jupiter \cite{Gierasch+2000}. The minimum NH$_3$ concentration is obtained for large values of $f_{\rm storm}$. The Juno MWR observations of a 100 to 250 ppmv ammonia abundance thus indicate that, at mid-latitudes, $f_{\rm storm}\ge 1$ (for $\epsilon=0.3$). On the contrary, the Equatorial Zone, represented by the right column of Fig.~\ref{fig:mush_model} is characterized by a relatively uniform ammonia abundance and thus requires $f_{\rm storm}\le 0.2$, in line with the lack of storms and lightning there. 

We \remove{can} thus \remove{explain}\add{obtain} a low abundance of ammonia to great depth \remove{if}\add{when} (1) strong storms are able to loft water ice particles into the mushball-formation region and (2) \remove{they occur more frequently than material is mixed upward in deep regions of Jupiter}\add{they involve, per unit time, more mass than does convection in deep regions of Jupiter}. This is a situation that appears to occur in most regions of Jupiter. In the Equatorial Zone these two conditions appear not to be fulfilled, explaining the high and relatively vertically uniform abundance of ammonia there. 

The temperature structure that can be inferred from Fig.~\ref{fig:mush_model} is characterized by a standard moist adiabatic profile in the Equatorial Zone\add{, in agreement with the analysis of }\cite{Li+2020}, and an extended moist adiabat driven by the evaporation of mushballs at mid-latitudes. Superadiabaticity factors may also play a role: while for Fig.~\ref{fig:mush_model} we assumed that $F_{\rm tot} /\dot{m}_{\rm deep} \ll wL_{\rm v}$, it may not be the case. In fact, in order to explain values $f_{\rm storm}>1$, the superadiabaticity at deep levels $\delta\theta_{\rm deep}$  should be larger than in the water condensation region $\delta\theta_{\rm storm}$, since in the absence of significant radiative transport, energy balance requires that $\dot{m}_{\rm deep} c_P \delta\theta_{/rm deep} \sim f_{\rm storm} \dot{m}_{\rm deep} c_P \delta\theta_{\rm storm}$ . This could lead to significant modifications of the interior adiabat and deserves detailed studies.

\subsection{Analytical solutions}

The system of equations defined by Eq.~\ref{eq:5layers} may be solved analytically with a few simplifications. First, we neglect the return upward flow arising from the fall of mushballs and evaporative downdrafts. This is justified as long as little atmospheric gas is entrained with mushballs and downdrafts (i.e., $\varpi_{\rm mush}\ll w_{\rm mush}/w$ and $\varpi_{\rm down}\ll w_{\rm mush}/w$). We then assume that water is abundant so that the mushball production is always limited by the availability of ammonia, i.e. that  \( w_{3}> \left( c_{2}-c_{\rm min} \right) f_{\rm H_2O}/f_{\rm NH_3} \). Finally, we ignore small-scale convection in the upper atmosphere (\( f_{\rm conv}=0 \)). In that case the system of equations yields:
\begin{equation}
\left\{
\begin{array}{l}
\displaystyle c_{1}=c_{2}= c_{\rm min}+\frac{c-c_{\rm min}}{1+ \epsilon +2 \epsilon f_{\rm storm}}\\
\displaystyle c_{3}= c_{\rm min}+\frac{ \left( c-c_{\rm min} \right)  \left( 1+ \epsilon  \right) }{1+ \epsilon +2 \epsilon f_{\rm storm}}\\
\displaystyle c_{4}= \frac{c \left( 1+ \epsilon + \epsilon f_{\rm storm} \right) +c_{\rm min} \epsilon f_{\rm storm}}{1+ \epsilon +2 \epsilon f_{\rm storm}}\\
\displaystyle w_{3}=w- \left( 2-a \right) \frac{ \left( c-c_{\rm min} \right)  \left( f_{\rm H_2O}/f_{\rm NH_3} \right)  \epsilon f_{\rm storm}}{1+ \epsilon +2 \epsilon f_{\rm storm}}\\
\displaystyle w_{4}=w- \left( 1-a \right) \frac{ \left( c-c_{\rm min} \right)  \left( f_{\rm H_2O}/f_{\rm NH_3} \right)  \epsilon f_{\rm storm}}{1+ \epsilon +2 \epsilon f_{\rm storm}}\\
\displaystyle s_{1}= s_{2}=s_0\\
\displaystyle s_{3}=s_0-L_{\rm v}w_{3}+\frac{F_{\rm tot}}{\dot{m}_{\rm deep} f_{\rm storm}}\\
\displaystyle s_{4}=s_0-L_{\rm v}w_{4}+\frac{F_{\rm tot} \left( 1+f_{\rm storm} \right) }{\dot{m}_{\rm deep} f_{\rm storm}}\\
\displaystyle s_{5}=s_0-L_{\rm v}w+\frac{F_{\rm tot} \left( 2+f_{\rm storm} \right) }{\dot{m}_{\rm deep} f_{\rm storm}}\\
\end{array}
\right.
\end{equation}

Thus when  \(  \epsilon f_{\rm storm} \gg 1 \),  \( c_{1}=c_{2}=c_{3} \approx c_{\rm min} \),  \( c_{4} \approx  \left( c+c_{\rm min} \right) /2 \)  and  \( w_{3} \approx w- \left( 1-a/2 \right)  \left( c-c_{\rm min} \right)  \left( f_{\rm H_2O}/f_{\rm NH_3} \right)  \),  \( w_{4} \approx w- \left( 1/2-a/2 \right)  \left( c-c_{\rm min} \right)  \left( f_{\rm H_2O}/f_{\rm NH_3} \right)  \) . When storms dominate the mass transport over the deep convection, the atmosphere is depleted in ammonia all the way to the deepest layer. The water abundance in layers 3 and 4 is controlled by the parameter  \( f_{\rm H_2O}/f_{\rm NH_3} \), i.e., by the ratio of water to ammonia in mushballs. This parameter crucially depends on the microphysics of particle growth and is thus very difficult to estimate, implying that we cannot at this point provide a quantitative estimate of the abundance of water. Importantly, in that limit, the process is independent of  \(  \epsilon  \), the efficiency of mushball formation.

The conditions for the mushball mechanism to deplete the deep atmosphere in ammonia can be derived from our analytical relations in the limit of negligible small-scale convection. A first condition is that mushball production should be limited by the availability of ammonia rather than water. This occurs when  \( f_{\rm NH_3}/f_{\rm H_2O}> \left(  c-c_{\rm min} \right) /w \) , implying  \( f_{\rm NH_3}\gtrsim0.09 \)  for a solar deep N/O ratio. The second condition is that  \( f_{\rm storm}\gtrsim1/  \epsilon  \) . Thus, even an inefficient mushball formation mechanism can lead to a depletion of ammonia to great depth, as long as storms are much more frequent than updrafts in the deep atmosphere, below the water cloud base. 

Since we are neglecting radiative heating and cooling, static energy is uniform in layers 1 and 2, a consequence of dry adiabatic motions by compensating subsidence. In the layers below, static energy decreases due to the evaporation of water ice and rain: the temperature gradient becomes smaller than a dry adiabat, and in fact equivalent to a moist adiabat. However it is important to note that this change extends even deeper than the water cloud base because of the sinking of mushballs to great depth. 

With these solutions, we can relate ammonia abundances (as found from MWR) to the value of the \( f_{\rm storm} \)  parameter. In order to consider both the equatorial region and the other latitudes, this time, we assume \( f_{\rm conv}=1 \). The relation between \( f_{\rm storm} \)  and  \( c_{3} \) is:
\begin{equation}
 f_{\rm storm}=\frac{c+c \epsilon -c_{3}-3 \epsilon c_{3}+2 \epsilon c_{\rm min}+\sqrt[]{8 \epsilon  \left( c-c_{3} \right)  \left( c_{3}-c_{\rm min} \right) + \left( c+c \epsilon - \left( 1+3 \epsilon  \right) c_{3}+2 \epsilon c_{\rm min} \right) ^{2}}}{4 \epsilon  \left( c_{3}-c_{\rm min} \right) } 
\label{eq:fstorm}
\end{equation}
This relation assumes  \( f_{\rm conv}=1 \), an approximation that allows to consider the equator and mid-latitude regions with the same model.

\section{Application to the MWR Juno results}\label{sec:applications}

\subsection{Reproducing the MWR Juno measurements}

We now compare the MWR ammonia abundance-latitude map to our theoretical model. In order to estimate the value of $f_{\rm storm}$ per latitude, we use Eq.~(\ref{eq:fstorm}) with the ammonia abundance from MWR (see Fig.~\ref{fig:model5layers}) in the 1-3 bar region. We then use this value in our full model defined by Eq.~(\ref{eq:5layers}) and our fiducial parameters from Table~\ref{tab:parameters}. We interpolate linearly the values of the mixing ratios as a function of depth (in $\log P$) to produce a map of the ammonia mixing ratios as a function of latitude and depth. 

\begin{figure}
\noindent\includegraphics[width=12cm]{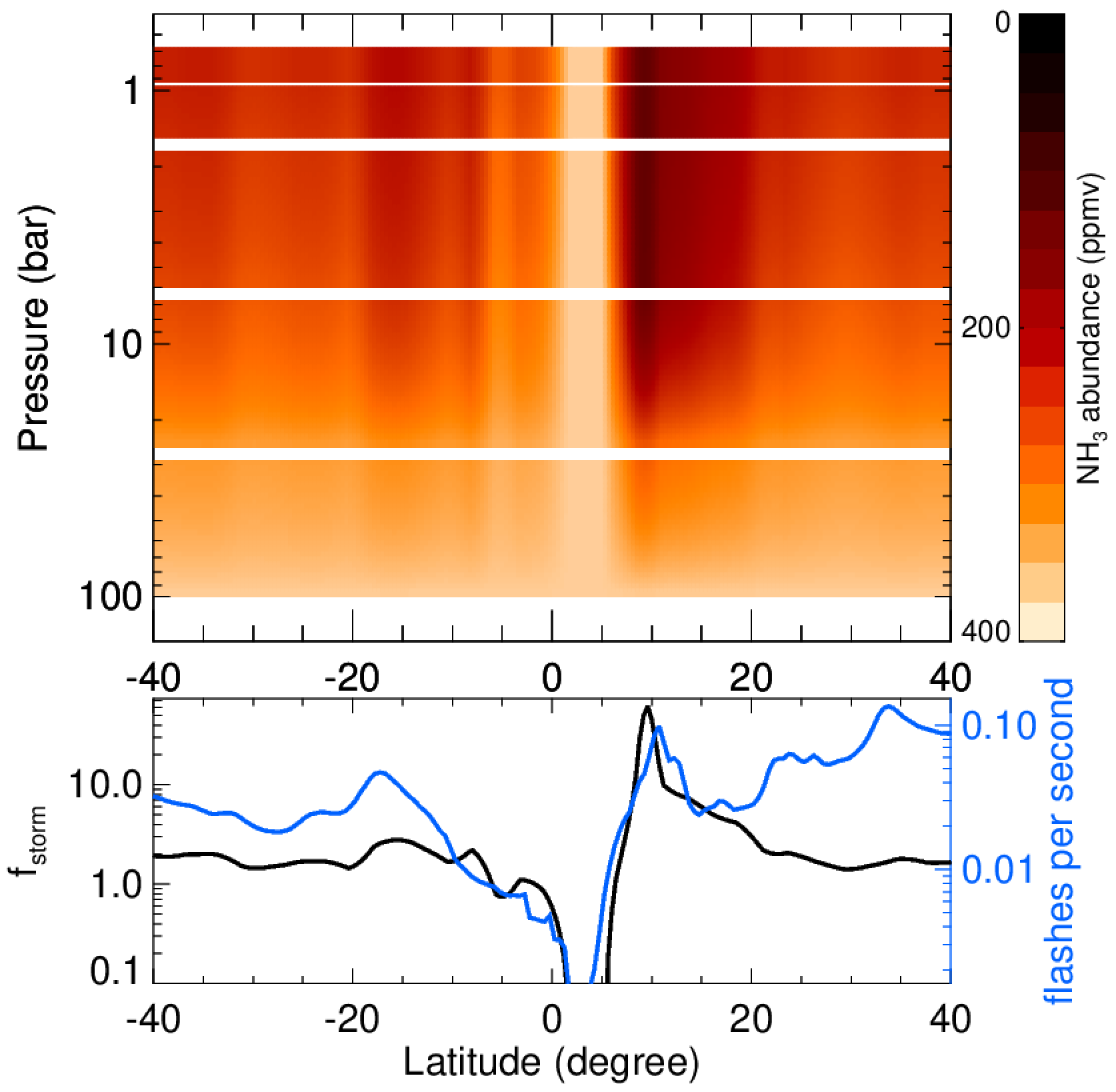}
\caption{Distribution of ammonia concentration obtained with our \remove{conceptual}\add{mass-balance} model. The top panel shows the NH$_3$ mixing ratio as a function of latitude and pressure in the 5 layers of our model. The bottom panel indicates the value of \add{$f_{\rm storm}$} (black line) obtained to reproduce the 1-3 bar MWR ammonia mixing ratio compared to the number of flashes per second detected by the MWR instrument between PJ1 and PJ16 \cite{Brown+2018}. The large and uniform ammonia concentration in the Equatorial Zone is well reproduced by assuming a scarcity of storms ($f_{\rm storm}\sim 0$), in line with the absence of lightning there. At mid-latitudes, frequent storms and subsequent mushball formation lead to a depletion of ammonia.  }
\label{fig:reproduce_MWR_map}
\end{figure}

The results are presented in Figure~\ref{fig:reproduce_MWR_map}. The dominant features, i.e., the nearly uniform abundance of ammonia in the Equatorial Zone and its depletion elsewhere \remove{are}\add{can be} explained by a change of the nature of convection at these latitudes, from being mostly small-scale (vertically) at the equator to being large-scale and dominated by water storms elsewhere. While our simple model is insufficient to explain the details of the ammonia distribution in the deep atmosphere, the suppression of storms at the equator is fully consistent with the Juno observation of a lack of lightning events \remove{at the equator} \cite{Brown+2018}, with the value of $f_{\rm storm}$ showing a clear correlation to the MWR lightning rate there (Fig~\ref{fig:reproduce_MWR_map}, bottom panel). The reason for the absence of storms itself, however, is not clear. It could be that vertical shear is absent in the Equatorial Zone and that the formation of rain and subsequent mass loading of water storms prevents their ascent \cite{Rafkin+Barth2015}. At the other extreme, it could also be that the Equatorial Zone experiences a very strong vertical shear that effectively extinguishes storm formation. Insolation, which is strongest at the equator, is also an important factor to consider \add{as it may bring the temperature profile at relatively low pressures ($P\lesssim 3$\,bar) slightly closer to an isotherm and suppress convection locally}.

\subsection{Ammonia and water}

The depletion of ammonia to great depths measured by Juno MWR is reminiscent of a long-standing issue, that of Jupiter's deep water abundance. Already in the 1980s, 5-$\mu$m spectroscopic observations of Jupiter's atmosphere had revealed a very low abundance of water vapor, one to two orders of magnitude less than the solar value, down to at least 6 bars in a wide region covering $-40^\circ$ to $+40^\circ$ latitude, with three times lower abundance in Jupiter's hot spots \cite{Bjoraker+1986}. A simple explanation was proposed: Jupiter's water clouds form narrow columns of humid air inside which water efficiently rains out to the cloud base, leaving the remaining region dry because of compensating subsidence \cite{Lunine+Hunten1987}. However this simple idea was shown to be incompatible with an Earth-based parametrization of cumulus clouds \cite{DelGenio+McGrattan1990}, for at least two reasons. First, compensating subsidence stabilizes the atmosphere and prevents further cumulus cloud activity, and second, upward mixing tends to bring moisture up from the cloud base level which is itself soaked by rain reevaporation. The picture, further strengthened by later detailed microphysical models \cite{Palotai+Dowling2008}, held to this day. When the Galileo probe measured an extremely low abundance of water in a 5-$\mu$m hot spot \cite{Niemann+1998, Wong+2004}, the explanation was that this was a special region of Jupiter, mostly downwelling and consequently dry, due to global-scale wave activity \cite{Ortiz+1998, Showman+Ingersoll1998, Showman+Dowling2000, Friedson2005}.

Yet, to this day, Jupiter's atmospheric water and ammonia abundances \add{calculated by} cloud models and global circulation models \cite{DelGenio+McGrattan1990, Palotai+Dowling2008} remain incompatible with retrievals from spectroscopic observations \remove{.}\add{:} The analysis of Galileo/NIMS and Juno/JIRAM spectroscopic observations \cite{Roos-Serote+2004, Grassi+2017, Grassi+2020} essentially confirm the previous observations by \citeA{Bjoraker+1986}. In order to reproduce the 5-$\mu$m spectra in the North Equatorial Belt, one generally requires a very low water abundance to great depths (8 bars or so), or at least a low relative humidity ($\sim 10\%$) until a cloud deck with a high opacity is reached. In addition, even though wave activity can explain qualitatively the low water abundance in 5-$\mu$m hot spots, the fact that the depletion persists down to at least 22 bars as measured by the Galileo probe remains unaccounted for. 

Our model accounts for a low ammonia abundance in region where storms are frequent. Because the fate of water is tied to that of ammonia, as shown in Fig.~\ref{fig:mush_model}, water is expected to be depleted as well. This could thus potentially explain the observations of both ammonia and water in Jupiter. The fact that this was not identified in previous studies is tied to three factors: (i) Hail is a very rare process on Earth and had always been neglected in studies of Jupiter's storms and general-circulation models. As shown in Paper~I, the presence of a region where a liquid $\rm NH_3\cdot H_2O$ mixture is bound to form is a pathway to hail formation. Such a property had not been identified previously, and thus hail formation was not considered in microphysical models \cite{Yair+1995, Palotai+Dowling2008, Sugiyama+2014, Li+Chen2019}. (ii) Evaporative downdrafts have small-scales and are notoriously difficult to model. As shown in Paper~I, they can efficiently transport a heavy condensable species even through layers where equilibrium chemistry would predict a complete mixing. (iii) Vertical diffusion by other processes was assumed to be more important than small-scale transport.

\subsection{Consequences for Jupiter's deep atmospheric structure}

Our model is bound to have strong consequences for Jupiter's deep atmospheric structure, in relation to its deeper internal structure. The molecular weight increase below the water condensation level due to the increase in both ammonia and water abundance is estimated to be of order $\Delta\mu/\mu\sim 10^{-2}$. (This is an order-of-magnitude value based on Fig.~\ref{fig:mush_model}, with our hypothesis of a solar N/O ratio). Because this takes place in a region where condensation is not possible, convection will be suppressed by this molecular weight gradient except where temperature fluctuations (or the temperature increase over a dry adiabat) is of order $\Delta T_\mu/T\sim \Delta\mu/\mu$, corresponding to a $3$\,K temperature increase at 300\,K. What seems like a tiny increase is in fact highly significant as can be seen from two quantities. 

First, let us introduce the convective available potential energy (CAPE) in the water-condensation region, which measures the \remove{ability of storms to develop and be extremely significant}\add{potential strength of storms, should they form} \cite{Holton1992}. The maximum value of this quantity can be calculated by assuming that the atmosphere follows a dry adiabat in the water-condensation region and that the humidity is 100\% at cloud base. In that case, the maximum energy released is approximately
\begin{equation}
{\rm CAPE_{Max}}=x_{\rm H_2O}(\mu_{\rm H_2O}/\mu)L_{\rm H_2O}\approx 46\times 10^7\ \rm erg/g, 
\end{equation}
for our fiducial water abundance (this value is of course proportional to the water abundance). Of course, this base temperature profile is violently unstable so that we expect in real situation much smaller values arising from a temperature gradient in the atmosphere that is close to a moist adiabat. On Earth, \remove{this value}\add{the value of ${\rm CAPE_{Max}}$} is similar (the mean molecular weight of the atmosphere is one order of magnitude higher, but 300\,K is reached near 1\,bar rather than near 6\,bar in Jupiter, implying that the water volume mixing ratio is about 6 times larger on Earth), but in fact the most violent thunderstorms generally associated with hail formation \remove{in the Earth atmosphere} occur when the value of CAPE reaches only about $5\times 10^7\,\rm erg/g$. 

In Jupiter, \add{we must also consider that} the increased temperature needed for a convective perturbation to bypass the molecular weight gradient is equivalent to an added CAPE 
\begin{equation}
\Delta {\rm CAPE}_\mu = c_{P,\,\rm atm} \Delta T_\mu \approx 85\times 10^7\ \rm erg/g,
\end{equation}
where we used $c_{P,\,\rm atm}=28\times 10^7\ \rm erg/(g\,K)$ (see Paper~I) and as above $\Delta T_\mu\approx 3\,$K.  Thus, deep convective events can potentially power extremely violent storms on Jupiter. Whether this is actually the case will depend on other processes, such as the balance between cooling by downdrafts and heating by convection from deeper regions. 

Another aspect to consider is the superadiabatic gradient needed to overcome the molecular weight gradient, i.e.,  $\nabla_{\rm s.ad}\equiv (d\ln T/d\ln P)-(\partial\ln T/\partial\ln P)_S\approx \Delta T_\mu /T /\Delta\ln P$, where $\Delta\ln P$ corresponds to the extent of the inhomogeneous region. Even if we consider that the region is extremely extended (say $\Delta \ln P=10$), this would imply a superadiabatic gradient $\nabla_{\rm s.ad} \gtrsim 10^{-3}$. In general, mixing length theory predicts that the superadiabatic gradient should be much smaller, i.e., $\nabla_{\rm s.ad} \lesssim 10^{-5}$ \cite{Guillot+2004}. This implies that convective events are transporting much more energy at a time and therefore should be much less frequent. Equivalently, this implies that the $\dot{m}_{\rm deep}$ parameter should be small, justifying a posteriori our finding that $f_{\rm storm}$ can be significantly larger than unity. 

Finally, it is important to note that evaporative downdrafts are delivering cool air to the deep atmosphere, providing another pathway to transport the internal heat from the deep region. This can potentially suppress convection at depth, in the downdraft region. For this to occur, the mushball flux needs to be such that the evaporative cooling balances the internal heating, i.e., $\dot{\tilde{m}}_{\rm mush}=F_{\rm tot}/L_{\rm v}\approx 3\times 10^{-7}\rm\,g/(cm^2\,s)$. In Jupiter, the number of storms per area is variable, but for example in the north equatorial belt it reaches $N_{\rm storms}\sim 2\times 10^{-9}\,\rm km^{-2}$ \cite{Brown+2018}. This implies that to offset convection at depth, each storm should dump $(F_{\rm tot}/L_{\rm v})/N_{\rm storms}\sim 1.5\times 10^{12}\,$g/s of condensates (mushballs). Assuming a typical storm area $\sigma_{\rm storm}\sim 300\rm\,km\,\times\,300\,km$, we can calculate that the precipitation rate should be $F_{\rm tot}/(L_{\rm v}\,N_{\rm storms}\,\tilde{\rho}\,\sigma_{\rm storm})\sim 6$\,cm/hr. On Earth, this would be classified as violent precipitation (in the form of rain, generally). With larger storm areas, an even weaker precipitation rate can offset heating by the planet's internal heat flux. 

This precipitation rate is significantly smaller than the maximum precipitation rate on Jupiter, obtained from $w \rho_{\rm cloud\ base} v_{\rm updraft} / \tilde{\rho}\sim 220$\,cm/hr. (We have assumed $w\sim 0.02$, $\rho_{\rm cloud\ base}\sim 5\times 10^{-3}\,\rm g/cm^3$, $v_{\rm updraft}=50\,$m/s). So even with an efficiency of $3\%$, strong storms in Jupiter may suppress convection at depth, providing a self-consistent explanation for the high $f_{\rm storm}$ values that we obtain at some latitudes.

\subsection{Caveats}

Of course, some important caveats must be added. We have neglected three crucial ingredients that eventually must be included: (i) large-scale advection and diffusion processes, (ii) radiative heating/cooling, and (iii) rotation. 

In our model, the ammonia (and water) transported downward by mushballs and evaporative downdrafts are only carried upward again by compensating subsidence. In the limit $f_{\rm storm}\gg 1$, this represents an absolute minimum to the amount of vertical transport and allows vertical abundance gradients to develop. Of course, observations of anticyclones and the relative success in modeling them \cite{Garcia-Melendo+2009,Palotai+2014} show that global-scale circulation matters. The MWR map from Fig.~\ref{fig:model5layers} show some structures that are not matched by our simple model in Fig.~\ref{fig:reproduce_MWR_map}. In reality, \remove{both} small-scale storms and large-scale circulation \remove{must play a role and shape the vortices that we see everywhere in Jupiter's atmosphere}\add{are interdependent and must both be considered to explain Jupiter's meteorology}. 

We have neglected radiative heating/cooling, and the frequency of storms that we infer is not self-consistently calculated as a function of stability arguments. We thus have not proven that we can self-consistently obtain high values of $f_{\rm storm}$ while transporting Jupiter's heat flux. This will require dedicated calculations including small-scale features such individual storms and large-scale structures with radiative transfer. The fact that the solar heating is strongly latitude-dependent yet measured atmospheric temperatures are nearly uniform \cite{Ingersoll+Porco1978} will have to be accounted for.  

Our model does not include rotation, which is certainly crucial to understanding the particularities of Jupiter's Equatorial Zone, i.e., the absence of strong storms and relative vertical uniformity of its ammonia abundance. We propose that the lack of storms at the equator may be related to \add{vertical} shear, but a quantified, predictive explanation is still lacking.

Finally, with only 5 layers, our model is extremely simplified and ignores important details. Our treatment of mixing small-scale convection imposes an arbitrary length-scale, i.e., the depth of each layer, when this should be treated as a diffusion equation with the proper parameters. The values of the $f_{\rm storm}$ parameter that we calculate are therefore only indicative and should not be used to quantify the strength of deep convection. We do not have enough resolution to distinguish between small water storms (which do not reach the 1.5-bar level) and large ones, implying that small water storms are treated as small-scale convection. This should not affect our results except quantitatively. We do not include other species, such as NH$_4$SH, which condenses around 2 bars and could sequester some of the nitrogen. Again, this should be marginal, owing to the small abundance of sulfur with respect to nitrogen in a solar-composition mix (i.e. S/N=0.19 according to \citeA{Lodders2003}).

\section{Conclusions}

We have shown that the variability of ammonia abundance in Jupiter retrieved by the Juno spacecraft \cite{Bolton+2017, Li+2017} can be linked to the presence of storms powered by water condensation. In paper I, we showed that powerful storms could deliver water ice particles to the 1.1-1.5 bar region where they would interact to form a liquid $\rm NH_3\cdot H_2O$ mixture that would lead to the formation of mushballs and evaporative downdrafts, potentially transporting ammonia to great depth. In the present paper, we developed a local model of Jupiter's deep atmosphere solving mass and energy balance to determine whether and in which conditions we could explain the Juno observations. 

Our model can account at least qualitatively for the observed vertical and latitudinal structure of the ammonia abundance in Jupiter. Storms powered by water condensation lead to the formation of mushballs and evaporative downdrafts and thus deplete the atmosphere of its ammonia and water locally. We introduced a parameter $f_{\rm storm}$, the ratio of the mass transported in these water storms to the mass transported by dry convection at greater depth, which measures the efficiency of the process. When $f_{\rm storm}\lesssim 1$, the process is inefficient and the ammonia abundance remains high. This is the situation corresponding to Jupiter's Equatorial Zone which is characterized by a high ammonia abundance \cite{Li+2017} and an absence of lightning flashes \cite{Brown+2018}. When $f_{\rm storm}\gg 1$, storms are dominating the mass transport, ammonia (and water) can be transported to great depth which explains the low mixing ratio of ammonia observed at all latitudes away from the $0^\circ-5^\circ$N region. When estimating the value of $f_{\rm storm}$ needed to reproduce the Juno ammonia measurements, we find that they are correlated to the flash rates measured by MWR, at least in the $-10^\circ$ to $10^\circ$ latitude range. Also, we find that at all latitudes, local maxima in $f_{\rm storm}$ correspond to local maxima of the flash rate. 

Importantly, the efficiency of the process results from a balance between the efficiency of mushball formation $\epsilon$ and the value of $f_{\rm storm}$. A low efficiency of mushball formation ($\epsilon\ll 1$) can lead to a significant depletion of ammonia with higher values of $f_{\rm storm}$. Of course important caveats, among them the fact that our model is purely local, that we did not consider radiative heat transport and that convective events are prescribed rather than self-consistently determined mean that this mechanism will have to be tested within cloud-ensemble models and general circulation models. 

Our model has a number of important consequences for Jupiter's deep atmosphere and interior: First, the equatorial region characterized by a well-mixed ammonia concentration, a lack of strong storms and of lightning flashes, should also be well-mixed in its water abundance. Its temperature structure is expected to be close to a standard moist-adiabat, \add{in agreement with the analysis of that region by} \citeA{Li+2020}. In contrast, we envision that the mid-latitude regions are not well-mixed in water, the increase in both water and ammonia abundance creating a region that is on average stably-stratified. The requirement to transport the internal heat flux implies that superadiabaticity should be significant, thus explaining, at least qualitatively, why $f_{\rm storm}$ can be significantly larger than unity. This \remove{can potentially have large implications to explain}\add{may have significant implications for} the internal structure of the planet \add{and can be tested by constraining the temperature profile in high-latitude regions. Additional measurements by the Juno spacecraft and combined analyses with ground-based data will be key to understanding the atmospheric variability and lifting the degeneracy between ammonia abundance and temperature.}

\add{Recently, the analysis of a powerful storm which occurred in Jupiter at latitude $16.5^\circ$S in January 2017 and was observed with multiple facilities including ALMA, VLA, HST (WFC2/UVIS), Gemini (NIRI), Keck (NIRSPEC), VLT (VISIR) and Subaru (COMICS) offers the possibility to test our model. This storm, which lasted about 3 weeks, led to an apparent increase of the ammonia abundance, reaching about 300\,ppmv} \cite{dePater+2019b}. \add{This may require adding an extra feature to our model, i.e., the presence of very deep plumes able to loft highly concentrated ammonia coming from the deepest regions. We note that the storm itself was complex, with both bright and dark features, probably indicating a combination of increase and decrease in ammonia concentration. Our model would accomodate a local increase in ammonia if there are also regions of low ammonia concentration (caused by mushball formation and evaporative downdrafts) so that on average, the ammonia abundance in the region} {\em decreases}. \add{Further study of this storm and similar ones is needed to validate our model.}   

The formation of mushballs and evaporative downdrafts should also occur in other giant planets in the solar system potentially explaining the low N/C ratio linked to the reported low ammonia abundances in the upper tropospheric region \cite{dePater+1991,Fletcher+2011,Irwin+2018,Guillot+Gautier2015}. The latitudinal distribution of ammonia in Saturn, although model-dependent and limited to the 1-3 bar region, appears to resemble that obtained for Jupiter with a peak in abundance at the equator and much lower values at mid-latitude \cite{Fletcher+2011}. The same study revealed that the tropospheric abundance of two disequilibrium species, arsine and phosphine, instead show a minimum at the equator, raising a conundrum \cite{Fletcher+2011}. This can now be understood in the framework of our model: strong storms, which are located away from the Equatorial Zone in mid-latitudes, deliver disequilibrium species from deep levels to elevate their abundance relative to the equator, but they tend to remove ammonia at mid-latitudes through the mushball process. 

Finally, we stress that the formation of mushballs lead to the presence of liquid (or partially liquid) condensates in a very high region of Jupiter's atmosphere that would otherwise contain only solids and vapor. 
The consequences of storms on the ammonia distribution may be observable by close-up MWR measurements from Juno \cite{Janssen+2017} over developing storms. The large-scale mid-latitude North Temperate Belt disturbances appear in Jupiter with a cadence of 4 years or so \cite{Sanchez-Lavega+2008, Sanchez-Lavega+2017} and would be an ideal candidate for an observation by Juno’s full set of instrumentation. \remove{Planets with hydrogen atmospheres remain crucial laboratories to understand atmospheric dynamics and meteorology in a regime in which condensates are heavier than the surrounding air Guillot1995.}

\acknowledgments
\add{This paper is dedicated to the memory of our friend and colleague Adam Showman, curious mind, great scientist and wonderful man. We thank the two reviewers for their careful reading of the manuscript and constructive comments.}  T.G. acknowledges support from the {\it Centre National d'Etudes Spatiales} and the Japan Society for the Promotion of Science. G.O. was supported by funds from NASA distributed to the Jet Propulsion Laboratory, California Institute of Technology. \remove{J.L. was supported by the Juno project through a subcontract from the Southwest Research Institute.}\add{A.I, J.L., P.S. and D.S. were
supported by NASA Contract NNM06AA75C from the Marshall Space Flight 
Center supporting the Juno Mission Science Team through a subcontract from the Southwest Research Institute.} 
The data used for this article is available at \remove{https://pubdata.space.swri.edu/look/0/7f50acda-30f9-43fe-b714-257c71d89404}\add{http://doi.org/10.5281/zenodo.3749573}


%
%

\bibliography{mushballs_jgr2}

%
%
%
%
%

\end{document}